\documentclass{svjour2}                  
\smartqed  
\usepackage{graphicx}
 \journalname{Foundation of Physics}
\begin{document}

\title{Is exponential metric a natural space-time metric of Newtonian gravity ?}


\author{M. Martinis  and
        N. Perkovi\'c 
}


\institute{Rudjer Bo\v{s}kovi\'{c} Institute \at
              Zagreb, Croatia\\
              Tel.: +385-1-4561032\\
              Fax: +385-1-4680223\\
              \email{martinis@irb.hr}
\\
}

\date{Received: date / Accepted: date}

\maketitle

\begin{abstract}
We show how Newtonian gravity with effective 
(actually observed) masses, obeying 
the mass-energy relation of special relativity, 
 can explain all observations used to 
test General relativity. Dynamics of a gravitationally coupled
binary system is considered   in detail, and the effective masses of  
constituents are determined.
 Interpreting our results in terms of  motion in a
curved space-time background,  we are led, 
using the  Lagrangian formalism,
 to consider  the exponential
 metric as a natural space-time metric  of Newtonian gravity.

\end{abstract}

\newpage
 
\section{Introduction}
\label{intro}

Newtonian dynamics introduced into scientific terminology the
concepts of inertial and gravitational mass, which for Newton were
mutually proportional quantities.The  exact  equality \cite{Dicke}
between inertial and gravitational mass was postulated by Einstein
\cite{Einstein} as the Principle of Equivalence upon which the
General Theory of Relativity was founded.
 Modern physical theories widely use two other  concepts of  mass,
  the invariant bare or proper mass  \cite{Okun}
  and the observer depended   effective mass \cite{Pinto}.
  The  effective  mass     can be viewed  as a  ''dressed''   bare mass  due to
   its interaction   with the   surrounding medium, the space-time background 
for example. 

  In this paper,  we reconsider the   motion of a point-like  objects through a 
gravitational background.  The interaction with the 
 background is described by the Newtonian-like gravity force with effective 
(actually observed) masses,
obeying the Einstein's mass-energy relation, $E = mc^2$ .
   The connection between bare and effective mass is then given by
   $m = m_0dt/d\tau  = m_gdt_g/d\tau$,where $t$ and $t_g$ are  the observer time and 
the gravitational time, respectively, while $\tau $ denotes
   the proper time of the moving object. The gravitational time is the time which shows
 the clock locally attached to the gravitational field.

   \section{Gravitational two body problem}

   In classical Newtonian mechanics the gravitational two body problem
   with inertial (bare) masses  is   exactly solvable  in an analytical form.
   However, these solutions fail  to explain    observed facts
   such as the     motion of planetary perihelion,
   the  starlight  deflection around the Sun, and the gravitational red shift.

   We shall show now that   Newtonian gravity with effective masses,
    obeying the mass-energy relation $E = mc^2$ can give us the
    satisfactory explanations to  all the  observations 
 used to test General relativity,
    without invoking the field equations of General    relativity.

    We begin by   considering   an isolated,
   gravitationally coupled,    binary system with point-like bare 
    masses $M_0$ and $m_0 << M_0$. In the rest frame  of $M_0$, the
    gravitational field
around $M_0$ is approximately static, spherically symmetric, and
isotropic.

In this  frame the orbital motion of  $m_0$   is
 described by the   Newtonian-like  gravity force 
\begin{equation}
 F_g =
 -GM_0m_g/r^2 = E_g\frac{du}{dr}
 \end{equation}
 where now $E_g = m_g c^2$ defines  the effective
 mass of $m_0$ which is actually observed from $M_0$.  
Since the  gravitational field
  around $M_0$ is static and spherically symmetric, the $m_g$ will depend only  
on the radial distance  from $M_0$.  Therefore, the $E_g$ changes according to
the well known rule  

\begin{equation}
 dE_g = drF_g = E_gdu,
 \end{equation}
  with  an obvious  solution
  \begin{equation}
  m_g = m_0e^u,
  \end{equation}
where $u = GM_0/rc^2 =R_g/r$ 
is the gravitational dimensionless potential of the mass $M_0$.
By observing  that the relativistic energy of $m_0$ can be written as  
$E = mc2 = m_gc^2/\sqrt{1 - \beta _g^2} = m_gc^2 dt_g/d\tau $,
we easily find the form of  $d\tau$ as
\begin{equation}
d\tau = dt_g\sqrt{1 - \beta _g^2} = dte^{-u}\sqrt{1 - e^{4u}\beta ^2},
\end{equation}                                                                                       [
where $\beta _g^2  = e^{4u}\beta ^2$ with $\beta ^2 = \vec{v}^2/c^2$.
Notice, that $dt_g = e^{-u}$ and $d\vec{r}_g = e^u d\vec{r}$ are 
time and space intervals measured in the coordinate frame  locally 
attached to the gravitational field.
The connection with the structure of the  space-time
 background  is obtained from  $ds = cd\tau =
 \sqrt{g_{\mu \nu}dx^{\mu}dx^{\nu}}$.  We find  that the 
background space-time metric is   isotropic and exponential: 
 \begin{equation}
 g_{00} = e^{- 2u}, \, \, g_{ii} = - e^{2u}, \, \, g_{0j} = 0.
\end {equation}
It is now easy to find the Lagrangian and the Hamiltonian  of the system.
From $Ldt = - m_0c ds$, it follows that
\begin{equation}
L  = - m_0c^2e^{-u}\sqrt{1 - e^{4u}\beta ^2},
\end{equation}
and from $H = v^ip_i - L$, we find the form of the Hamiltonian
\begin{equation}
H = e^{- u}\sqrt{p_ip^ic^2 + m_0^2c^4} = e^{-2u}E,
\end{equation}

 where $p_i = \frac{\partial L}{\partial v^i} = e^{2u}p^i$ denotes 
the canonical 3-momenta, and \\ $p^i =
v^iE/c^2$ with $v^i = dr^i/dt$.  

   The orbital motion of $m_0$  is best analyzed  in the  polar coordinate frame,
   ($r, \theta,\varphi$).There are  two constants
    of motion $H$ and $L$ which   are  obtained from the Lagrangian
  equation of motion. They  express  the fact that  the gravitational field is static ($H$) 
and sperically symmetric ($L$).   
 .
  The conservation of $L$ enable us to consider  the equatorial motion ,
   ( $ \theta = \pi/2$),of $m_0$ only, given by
     
\begin{equation}
    \Big(\frac{dr}{ds}\Big)^2 =
    (h^2 - e^{-2u}-    e^{-4u}\frac{l_{\varphi}^2}{r^2})
    \end{equation}

    where $h = H/m_0c^2$ and
    $ l_{\varphi} = L/m_0c = e^{2u}r^2d\varphi/ds$ 
are now those two constants     of motion.   
    
          In the  ($ u,\varphi$)- plane, 
    we have to solve two equivalent differential equations 
    \begin{equation}
    u'^2  =
    (h^2e^{2u} - 1)\frac{e^{2u}}{l^2} - u^2
    \end{equation}
      and
  \begin{equation}
u'' + u = \frac{e^{2u}}{l^2} + 2(u'^2 + u^2)
\end{equation}

where $u'= du/d\varphi$ and  $l = l_{\varphi}/R_g$.

  The exact analytic solution of these two equations is not yet feasible, but 
the required accuracy of an approximate solution can be achieved by expanding 
the right hand side  of $u'^2$ in a truncated power series  in the variable  $u\ll 1$:
\begin{equation}
u'^2  = 2[ 2\epsilon  + (4\epsilon + 1)u + (8\epsilon + 3)u^2 + ... ]/l^2 -  u^2.
\end{equation}
where $2\epsilon = h^2 - 1 < 0$ for a bound motion.

 In the region  where $|\epsilon |$ and $u$ are very small, 
a close graphical inspection of the right hand side of $u'^2$ shows that already 
a quadratic polynomial     is a very good approximation to $u'^2$ 
and the results we are going to derive.

The equations to be analyzed now are of the form
\begin{equation}
u'^2 + u^2 = \frac{2}{l^2}(\epsilon + u + 3u^2)
\end{equation}
and
\begin{equation}
u'' + (1 - \frac{6}{l^2})u = \frac{1}{l^2} .
\end{equation} 

 The classical (non-relativistic)  solution is obtained if
  the   relativistic term $(6/l^2)u$ is neglected in (13). 
 We are then left with the equation
\begin{equation}
 u'' + u = \frac{1}{l^2}
\end{equation}
 which has   the  standard  Newtonian solution
\begin{equation}
u(\varphi ) = \frac{1}{l^2} (1 + e \cos \varphi)
\end{equation}
This particular solution is designed so that  $u(0) =u_1 = u_{max}$ and \\
$u(\pi ) = u_2 = u_{min}$.
The    eccentricity $e$ of the elliptical orbit is $e < 1$,and is  given by
\begin{equation}
e = \frac{u_1 - u_2}{u_1 + u_2} = \sqrt{1 + 2\epsilon l^2}.
\end{equation}
We note that  the classical solution also obeys  the relation    
 $u'(0) = u'(\pi ) = 0$. This relation would not 
 be obeyed   if    relativistic corrections were included,
namely in this case $ u'(\pi ) \not= 0$.
 
    \section{ Motion of the  perihelion}

We show now  that   (13) for $e < 1$ describes  
an ellipse with moving perihelion.  
The solution of (13) can be writen in the following  form
\begin{equation}
u(\varphi )  = \frac{1}{l_r ^2}(1 + e_r \cos(\varphi \sqrt{1 - 6l^{-2}})) 
\end{equation}
where $l_r^2 =l^2 - 6$, and $e_r = \sqrt{1 + 2\epsilon l_r^2}$.
The  turning points of the orbit are  defined by $u'(\varphi ) = 0$.
They are   found  at 
$\varphi = \varphi _n = (n - 1)\pi /\sqrt{1 - 6l^{-2}}$, where  $n = 1, 2,3,...$.
The angle by which the perihelion of the elliptical orbit is shifted
in the direction of motion  per one revolution
is given by
\begin{equation}
\Delta \varphi  = 2(\varphi _{n+1} - \varphi _n - \pi) =
 2\pi (\frac{1}{\sqrt{1 - 6l^{-2}}} - 1) \approx \frac{6\pi }{l^2}.
\end{equation}
Here we can write, with sufficient accuracy, $l = (c/a\omega )\sqrt{1 - e^2}$,
where $a$ is the major semi-axis and $\omega = 2\pi /T$ with
$T$ being the period of revolution. The final and  well known expression 
for  the perihelion shift per period is obtained
\begin{equation}
 \Delta \varphi  = 24\pi ^3\frac{a^2}{c^2T^2(1 - e^2)}.
\end{equation}
In the classical non-relativistic limit  the motion of perihelion 
 is not predicted, and $\Delta \varphi _{classical} = 0$. 

Attempts to explain the motion of the planetary perihelion, using only 
special relativity and   modified by hand
inertial masses  in Newtonian gravity, date
 back to 1917 and earlier years \cite{Silver}.

\section{Gravitational light deflection}

The gravitational light deflection  around the Sun 
 was one of the crucial tests 
of General relativity in 1919. 
By allowing  the light pulse to have a mass that moves 
 with the velocity  of light, the calculations  using Newtonian gravity 
  gave  only half a value that was observed for the light deflection angle.   
\\ We show now, that  calculations
  using  the Newtonian gravity f with effective masses 
obeying the mass-energy relation would give the   observed 
 values for a light deflection angle.  
According to special relativity, light pulse has   a vanishing  bare (rest) mass. 
Light motion through the static and spherically symmetric gravitational field is 
described  by the  Hamiltonian, $H = c\sqrt{p^i p^i}$, which is obtained from (7) 
by setting $m_0 = 0$. In this limit ,$ l \rightarrow \infty $, but in such a way
that $h/l  \approx u_1$ remains  constant . The light trajectories are  then determined  by
solving the following two differential  equations:
 \begin{equation}
    u'^2 + u^2 =
    (\frac{h}{l})^2e^{4u} \approx u_1^2 [ 1 + 4(u - u_1) + 8(u - u_1)^2 + ... ]
    \end{equation}
      and
  \begin{equation} 
u'' + u = 2(u'^2 + u^2) \approx 2u_1^2 [ 1 + 4(u - u_1) + 8(u - u_1)^2 + ... ].
\end{equation}
where $u_1 = R_g/r_{min}$ and $r_{min}$ is the closest distance  that the 
light pulse gets to  the source of gravity field  $M_0$. 
The  appropriate coordinate frame 
 for studying shapes of  light trajectories is
 $u(\varphi) = u(- \varphi ), u(0) = u_1, u'(0) = 0$ and 
$u(\pi /2 + \Delta \varphi /2) = 0$. Here, $\Delta \varphi $
 is by definition   the    deflection
angle for light in a gravitational field. In the  weak field limit, $u\ll 1$, 
it is sufficient to study  the  equation
\begin{equation}
u''  + u = 2u_1^2.
\end{equation}
in order to determine the shape 
of  the light trajectory.  
In our coordinate frame,the  solution of (21)  is $ u(\varphi ) = 2u_1^2 + B\cos\varphi $, 
where $B = u_1(1 -2u_1)$. The condition $u(\pi/2 +\Delta \varphi /2) = 0$
yields to  $\sin(\Delta \varphi /2) = 2u_1/(1 - 2u_1)$ from which we get,
 to  lowest order in $\Delta \varphi $ and $u_1$, the final expression ,  
for the  deflecting angle of light in a gravitational field:
\begin{equation}
\Delta \varphi  = 4u_1 = \frac{4GM_0}{c^2r_{min}},
\end{equation}
This result is found by using Newtonian gravity with effective masses 
obeying the mass-energy relation. Classical calculations, using 
Newtonian gravity with bare masses, 
would lead    to the equation $u'' + u = u_1^2$ 
for the  light trajectory.  This equation predicts  
 a deflection angle $\Delta \varphi _{classical} = 2u_1$ 
that is half the value that was observed.

\section{Conclusion}

In this paper   we studied the implications of interpreting the 
Newtonian gravity force in terms of actually observed effective 
masses obeying the mass-energy relation of special relativity. 
We claimed that the gravitational attraction should be 
acting between all mass-equivalent energies. 
Using the   Lagrangian formalism, the corresponding space-time 
background metric is found  
to be exponential.  
This metric was first introduced by
Yilmaz \cite{Yilmaz}   in an attempt to modify the Einstein's field equations of General
relativity. However, the Yilmaz theory was immediately sharply 
criticized \cite{Misner}  on various
grounds \cite{Ibison}  as being ill de¯ned and because it does not predict 
 existence of  black-holes \cite{Robert}.
In our  approach, however,
  the exponential metric arises  as a consequence of introducing an
observer in the process of measuring the gravitational attraction
by means of the  Newtonian gravity with effective masses. We claim that   
the proper (bare) mass $m_0$
of a body  in a gravitational field will   always be 
observed as an effective mass $m_g = m_0e^u$.  
For example, on the Earth's surface,    the ratio $m_g/m_0$ 
  is very close to unity,
or more precisely $\frac{m_g}{m_0} - 1 \approx 7\times 10^{-10}$. 

The exponential metric belongs to a large family 
 of alternative theories of gravity, which all agree with General relativity 
to a first order in $u$.

In this paper, we have also demonstrated how  Newtonian   gravity  
with effective masses explains the motion of perihelion in binary 
systems,and  the gravitational deflection  of light rays.
The gravitational red-shift can also be explained 
by our theory \cite{Martin} from  the relation
\[ \frac{(E_g(2) - E_g(1)}{E_g(1)} = e^{u(2) - u(1)} - 1 \approx u(2) - u(1) + ... \]
This result agrees both with the prediction of General relativity and with observations.
 
\begin{acknowledgements}
This work was supported by the Croatian Ministry of Science, Education and
Sport, Project No.098-0982930-2900.
\end{acknowledgements}

  \end{document}